\documentclass[11pt]{article}
\usepackage{amsmath,amssymb}
\usepackage[margin=1in]{geometry}
\usepackage[skip=4pt]{parskip}
\usepackage{tikz}
\usepackage[numbers]{natbib}   
\usepackage{hyperref}
\usetikzlibrary{decorations.pathmorphing,patterns}

\newcommand{\zt}{\tilde z}
\newcommand{\yt}{\tilde y}

\title{The Schwitters Memorial Problem:\\
A Car Speeds Across a Triangular Bump}
\author{William H. Press\thanks{Corresponding Author: email wpress@utexas.edu}\\
Oden Institute for Computational Engineering and Sciences\\ 
The University of Texas at Austin\\}
\date{\today}

\begin{document}
\maketitle

\section{Introduction}

Parkway, in Austin, Texas, is a north-south street that runs along the west side of Pease Park. Savvy commuters use it to avoid heavy traffic on Lamar Boulevard, which runs along the east side of the park, but they, including myself and my late friend and colleague Roy Schwitters (1944-2023) \cite{Riordan2023,airhart2023,wikiroy}, are relatively few. Perhaps the reason is that Parkway has speed bumps (also known as speed humps or speed cushions) every few hundred meters along its length. These speed bumps span about 2 m along the direction of travel, rising at their crest by perhaps 15 cm. They appear to be made of some durable rubber composite.

Most drivers slow down for the speed bumps, sometimes to almost a stop---as if each speed bump were a stop or yield sign. Schwitters and I were long of the opinion that one should take them at speed, not less than 60 or 70 kph (35 to 45 mph), not because we were scofflaws, but because we thought, based on experiment, that the traverse was less jarring at high speeds than at low. Here, I will call this the Schwitters Doctrine: Take speed bumps safely, but, that said, {\em faster is better}.

We each expressed our intention to work through the elementary physics of what was clearly some kind of damped harmonic oscillator problem \cite{openstax,mathworks,young_freedman}.  Each urged the other to do so. But neither of us did. Roy died in Orcas Island, Washington, on 10 January, 2023. I regret that it has taken me so long to get around to this, especially now that the problem reveals unexpected richness that, hopefully, the reader will also enjoy.

\section{Statement of the problem}

A car drives at constant speed $v$ over an isosceles triangular bump of
height $h$ and half-width $w$. The wheel, of radius $r$, is connected
to the car body through a suspension (a spring and a damper in
parallel). \emph{What does the driver feel?} The driver's cushioned seat rides
with the car body. The physically felt quantity is the impulse and/or acceleration and/or
jerk (the next derivative) delivered to the driver's seated rear.
The deliverable of this problem is therefore the
body motion $z(t)$ and an interpretation of how it is perceived by the driver, as a function of the crossing speed $v$.

The analysis proceeds in stages:
a purely geometric calculation of the axle path (\S\ref{sec:geometry});
a simple damped harmonic oscillator formulation (\S\ref{sec:oscillator});
nondimensionalization of the
equation, reducing its number of parameters and allowing an analytic solution
(\S\ref{sec:nondim});
allowing for the fact that the wheel can bounce off the road (\S\ref{sec:contact});
discussion of the model's limitations (\S\ref{sec:limitations});
and a review of related work on washboard roads (\S\ref{sec:related}).

Throughout, when calculating typical numerical values, we use the following values for parameters in cgs units. (The reader may substitute at will other values into the formulas.)
\begin{center}
\begin{tabular}{llll}
\hline
quantity & symbol & value & \\
\hline
wheel radius        & $r$       & $35$ cm & $\approx$ tire size 225/60R18\\
bump height         & $h$       & $15$ cm & \\
bump half-width     & $w$       & $100$ cm & \\
loaded suspension frequency & $f_0$ & $1$ Hz & (typical: 0.8--1.2 Hz)\\
damping ratio       & $\zeta$   & $0.3$   & (typical: 0.25--0.35)\\
gravity             & $g$       & $980$ cm/s$^2$ & \\
speed            & $v$       & $900$ cm/s & $32$ km/hr $\approx 20$ mph\\
sprung quarter-car mass & $M$ & --- & analysis independent of $M$ \\
unsprung wheel mass ratio & $m_w/M$ & 0.1 & used in \S7 only\\
\hline
\end{tabular}
\end{center}

\section{Geometry: the path of the axle}
\label{sec:geometry}

\subsection*{Exact result}

\begin{figure}[t]
\centering
\begin{tikzpicture}[scale=1.05]
  \draw[thick] (-5.3,0) -- (-3,0) -- (0,1.2) -- (3,0) -- (5.3,0);
  \draw[<->] (0,-0.45) -- (3,-0.45) node[midway,below]{$w$};
  \draw[densely dotted] (3,0) -- (3,-0.55);
  \draw[densely dotted] (0,1.2) -- (0,-0.55);
  \draw[densely dotted] (0,1.2) -- (4.6,1.2);
  \draw[<->] (4.6,0) -- (4.6,1.2) node[midway,right]{$h$};
  \draw[thick] (1.5,1.8924) circle (1.2);
  \fill (1.5,1.8924) circle (0.055);
  \draw (1.5,1.8924) -- (1.0543,0.7782) node[midway,below right]{$r$};
  \draw[red,dashed,very thick]
    (-5.3,1.2) -- (-3.2311,1.2) -- (-0.4457,2.3142)
    arc[start angle=111.82,end angle=68.18,radius=1.2]
    -- (3.2311,1.2) -- (5.3,1.2);
  \node[red] at (-4.4,1.45) {$y(x)$};
  \fill[red] (3.2311,1.2) circle (0.055) node[above right]{$P$};
  \fill[red] (0.4457,2.3142) circle (0.055) node[above right]{$Q$};
  \fill[red] (-3.2311,1.2) circle (0.055) node[above left]{$P'$};
  \fill[red] (-0.4457,2.3142) circle (0.055) node[above left]{$Q'$};
  \draw[densely dotted,red] (3.2311,1.2) -- (3.2311,0)
    node[below,black]{$x_{\mathrm P}$};
  \draw[densely dotted,red] (0.4457,2.3142) -- (0.4457,0)
    node[below,black]{$x_{\mathrm Q}$};
\end{tikzpicture}
\caption{Geometry of the axle path (red dashed) as the wheel crosses
the bump: flat ($y=r$), a line parallel to each face at offset $r$, and
an arc of radius $r$ about the apex, joined at $\pm x_{\mathrm P}$
(slope discontinuity) and $\pm x_{\mathrm Q}$ (tangent). Drawn with
$h/w = 0.4$, exaggerated for clarity; the analysis assumes
$h/w \ll 1$.}
\label{fig:geometry}
\end{figure}
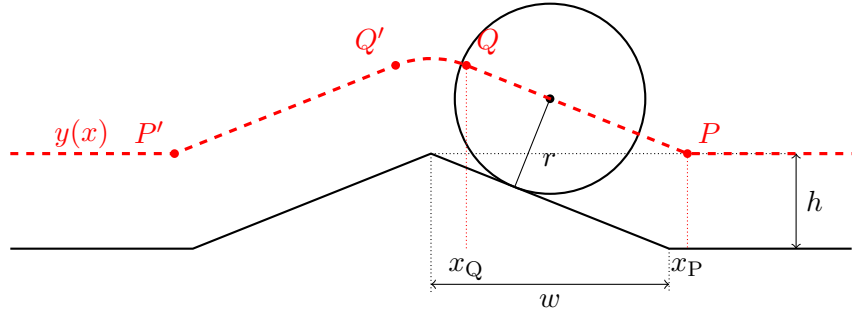

Place the origin at the foot of the apex, so the road surface is
\begin{equation}
  y_{\mathrm{s}}(x) =
  \begin{cases}
    h - (h/w)\,|x|, & |x| \le w,\\
    0, & |x| > w .
  \end{cases}
\end{equation}
We seek the height $y(x)$ of the wheel center (the axle) as a function
of its horizontal position $x$, assuming the rim maintains contact with
the surface. The motion has five phases (Figure~\ref{fig:geometry}).

On flat ground the contact point is directly below the axle and
$y = r$. While the rim rolls on a face of the bump, the axle travels on
a line parallel to that face at perpendicular distance $r$. Displacing
the right face $y = h - (h/w)x$ along its upward unit normal
$\hat n = (h,w)/\sqrt{h^2+w^2}$ raises the intercept by
$r\sqrt{1+(h/w)^2}$, giving
\begin{equation}
  y = h - \frac{h}{w}\,x + r\sqrt{1+(h/w)^{2}}
  \label{eq:offsetline}
\end{equation}
(and its mirror image on the left face). Near the top, the apex is a
\emph{convex} corner: the rim pivots about the single point $(0,h)$ and
the axle traces a circular arc of radius $r$,
\begin{equation}
  y = h + \sqrt{r^{2}-x^{2}} .
  \label{eq:arc}
\end{equation}

The transitions between phases occur at two points (and their mirror
images). The base corner $(w,0)$ is \emph{concave}, so the handoff from
flat ground to face contact is instantaneous, at the intersection of
$y = r$ with the offset line \eqref{eq:offsetline}:
\begin{equation}
  x_{\mathrm{P}} = w + r\,\frac{w}{h}
    \left(\sqrt{1+\frac{h^{2}}{w^{2}}}-1\right),
  \qquad y_{\mathrm{P}} = r .
  \label{eq:xP}
\end{equation}
Note $x_{\mathrm{P}} > w$: the axle begins to rise \emph{before}
passing the base corner, because the rim reaches forward and touches
the face early. The offset line \eqref{eq:offsetline} lies at
perpendicular distance exactly $r$ from the apex (the face passes
through the apex), so it is automatically \emph{tangent} to the arc
\eqref{eq:arc}; the tangent point is the apex displaced by $r\hat n$,
\begin{equation}
  x_{\mathrm{Q}} = \frac{r h}{\sqrt{h^{2}+w^{2}}},
  \qquad
  y_{\mathrm{Q}} = h + \frac{r w}{\sqrt{h^{2}+w^{2}}} .
  \label{eq:xQ}
\end{equation}
Both $y$ and $dy/dx$ are continuous at $x_{\mathrm{Q}}$; at
$x_{\mathrm{P}}$ only $y$ is continuous. Collecting the phases,
\begin{equation}
  y(x) =
  \begin{cases}
    r, & |x| > x_{\mathrm{P}},\\[2pt]
    h - \dfrac{h}{w}\,|x| + r\sqrt{1+(h/w)^{2}},
      & x_{\mathrm{Q}} < |x| \le x_{\mathrm{P}},\\[8pt]
    h + \sqrt{r^{2}-x^{2}}, & |x| \le x_{\mathrm{Q}} .
  \end{cases}
  \label{eq:piecewise}
\end{equation}
The construction requires
that the wheel be unable to bridge from flat ground directly onto the
apex; the condition works out to $rh < 2w^{2}$, comfortably satisfied
here ($525 \ll 20\,000$~cm$^2$).

\subsection*{The shallow-bump limit}
\label{sec2p2}

For $h/w \ll 1$ the wheel-radius corrections in \eqref{eq:piecewise}
are all small. The uniform lift of the face segments,
$r\bigl[\sqrt{1+(h/w)^2}-1\bigr] \approx rh^{2}/2w^{2}$, is negligible
compared to $h$; the onset corners shift outward by
$x_{\mathrm{P}} - w \approx rh/2w$ ($2.6$~cm here), absorbable into
the time origin; and the sharp apex is rounded over the small distance
\begin{equation}
  x_{\mathrm{Q}} \approx r\,\frac{h}{w} \;\ll\; r
\end{equation}
($5.25$~cm here), with a depth deficit $rh^{2}/2w^{2}$ below the true
apex. All three corrections are governed by the single small parameter
$rh/w^{2}$ (here $0.0525$). To leading order the axle simply retraces
the bump profile itself, lifted by $r$: the wheel acts as a low-pass
filter that smooths only features sharper than its own radius.

One feature of \eqref{eq:piecewise} survives at any $h/w$ and will
drive everything that follows: the axle path has \emph{kinks} --- slope
discontinuities at $\pm x_{\mathrm{P}}$, and (once the apex rounding is
neglected) at the apex as well. Swept at constant speed, $x = vt$,
these become jumps in the axle's vertical velocity, i.e.\ delta-function
vertical accelerations. How the suspension shields the driver from
them is the subject of the rest of the paper.

\section{The car as a damped oscillator}
\label{sec:oscillator}

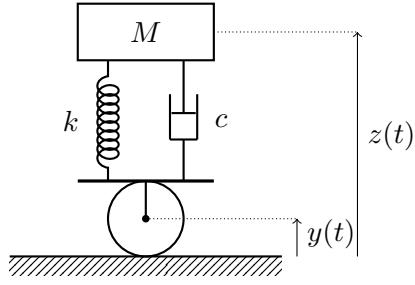
\begin{figure}[t]
\centering
\begin{tikzpicture}[scale=1.0]
  \fill[pattern=north east lines] (-1.8,-0.25) rectangle (1.8,0);
  \draw[thick] (-1.8,0) -- (1.8,0);
  \draw[thick] (0,0.5) circle (0.5);
  \fill (0,0.5) circle (0.05);
  \draw[very thick] (-0.9,1.0) -- (0.9,1.0);
  \draw[thick] (0,0.5) -- (0,1.0);
  \draw[thick,decorate,decoration={coil,aspect=0.7,segment length=3.4pt,
        amplitude=4pt,pre length=5pt,post length=5pt}]
    (-0.5,1.0) -- (-0.5,2.6);
  \node[left] at (-0.75,1.8) {$k$};
  \draw[thick] (0.5,1.0) -- (0.5,1.55);
  \draw[thick] (0.32,1.55) -- (0.32,2.15);
  \draw[thick] (0.68,1.55) -- (0.68,2.15);
  \draw[thick] (0.32,1.55) -- (0.68,1.55);
  \draw[thick] (0.5,2.6) -- (0.5,1.9);
  \draw[thick] (0.34,1.9) -- (0.66,1.9);
  \node[right] at (0.78,1.8) {$c$};
  \draw[thick] (-0.9,2.6) rectangle (0.9,3.35);
  \node at (0,2.975) {$M$};
  \draw[densely dotted] (0,0.5) -- (2.0,0.5);
  \draw[->] (2.0,0) -- (2.0,0.5);
  \node[right] at (2.0,0.3) {$y(t)$};
  \draw[densely dotted] (0.9,2.975) -- (2.8,2.975);
  \draw[->] (2.8,0) -- (2.8,2.975);
  \node[right] at (2.8,1.6) {$z(t)$};
\end{tikzpicture}
\caption{Quarter-car model: the body (sprung mass $M$) rides on the
axle through a damped spring with spring constant $k$ and damping constant $c$. The axle height
$y(t)$ is prescribed by the geometry of \S\ref{sec:geometry} with
$x = vt$; the body height $z(t)$ is the dynamical degree of freedom.
The arrows are schematic: in the equations both $y$ and $z$ are
measured from their static-equilibrium values.}
\label{fig:quartercar}
\end{figure}

Mount the car body, of (quarter-car) mass $M$, on the axle through a damped spring with spring constant $k$ and damping constant $c$
(Figure~\ref{fig:quartercar}). Collectively, the spring and damper (shock-absorber) comprise the strut or suspension that holds up a quarter of the car. The axle height $y$, given by
\eqref{eq:piecewise} with $x = vt$, is the prescribed base excitation;
the body height $z(t)$ is the unknown. Measuring both $z$ and $y$ from
their static-equilibrium values (the body from the static ride height,
where the spring compression $Mg/k$ balances the weight; the axle from
$y = r$), gravity and the spring's natural length cancel identically,
and Newton's second law for the body reads
\begin{equation}
  M\ddot z + c\,(\dot z - \dot y) + k\,(z - y) = 0 .
  \label{eq:eom-dim}
\end{equation}
Only relative displacements and velocities of the two ends of the
suspension appear, which is why the constant offsets are harmless.
(Gravity has not disappeared from the problem: it reappears in
\S\ref{sec:contact}, where the \emph{absolute} normal force at the
road matters.)

Define the loaded natural frequency and damping ratio in the standard
way,
\begin{equation}
  \omega_0 = 2\pi f_0 = \sqrt{\frac{k}{M}},
  \qquad
  \zeta = \frac{c}{2M\omega_0} = \frac{c}{2\sqrt{kM}} ,
  \label{eq:defs}
\end{equation}
``loaded'' emphasizing that $M$ is the mass actually riding on the
spring, so $f_0$ is the frequency at which the laden car bounces.
Dividing \eqref{eq:eom-dim} by $M$:
\begin{equation}
  \ddot z + 2\zeta\omega_0\,(\dot z - \dot y)
          + \omega_0^{2}\,(z - y) = 0 .
  \label{eq:eom-omega}
\end{equation}

\section{Nondimensionalization and Analytic Solution}
\label{sec:nondim}

We will now see that equation \eqref{eq:eom-omega} can be solved exactly analytically in the shallow-bump limit of \S\ref{sec2p2}, where the axle displacement (measured from $y = r$) is the triangular pulse itself.

It is useful to first nondimensionalize the ODE. In doing so, we will discover that the problem has essentially only one free parameter--the bump crossing time in units of the car's suspension period. (While $\zeta$, the damping ratio, is technically another free parameter, its value for real vehicles is never far from 0.3.)

\subsection*{Nondimensionalization}
Choose $t = 0$ at the apex and let
\begin{equation}
  \tau_0 \equiv \frac{w}{v}
\end{equation}
be the \emph{half-crossing time} (base corner to apex). Measure time in
units of $1/f_0$ and displacements in units of the bump height $h$
(going directly to the final scaling; an intermediate scaling by $r$
would introduce $h/r$ as a third parameter, but since the ODE is linear
and homogeneous in displacement, $h/r$ is a pure amplitude factor and
scales out):
\begin{equation}
  \tau = f_0\, t,
  \qquad
  \zt = \frac{z}{h},
  \qquad
  \yt = \frac{y - r}{h},
  \qquad
  T \equiv f_0\,\tau_0 = \frac{f_0\, w}{v} .
  \label{eq:scalings}
\end{equation}
Here $T$ is the half-crossing time in units of the suspension period.
Each time derivative contributes a factor $f_0$; dividing
\eqref{eq:eom-omega} by $h f_0^{2}$ gives
\begin{equation}
  \zt'' + 4\pi\zeta\,(\zt' - \yt\,')
        + (2\pi)^{2}\,(\zt - \yt) = 0 ,
  \qquad
  \yt(\tau) = \Lambda\!\left(\frac{\tau}{T}\right),
  \qquad
  \Lambda(s) \equiv \max(0,\,1-|s|),
  \label{eq:eom-nondim}
\end{equation}
with primes denoting $d/d\tau$ and $\omega_0 = 2\pi$ in these units.
The response \emph{shape} is governed by just two dimensionless
parameters,
\begin{equation}
  T = \frac{f_0 w}{v}
  \qquad\text{and}\qquad
  \zeta ,
\end{equation}
with an interesting regime evidently $T \sim 1$, where the bump
crossing resonates with the suspension. The bump amplitude $h$ (or
$h/r$, if $r$ is preferred as the length unit) multiplies the entire
solution but does not affect its form.

To return to physical variables:
\begin{equation}
  t = \frac{\tau}{f_0},
  \qquad
  z = h\,\zt,
  \qquad
  \dot z = h f_0\,\zt',
  \qquad
  \ddot z = h f_0^{2}\,\zt'' .
  \label{eq:inverse}
\end{equation}

The forcing $\yt$ is continuous but kinked at the three instants
$\tau = -T,\ 0,\ +T$: its derivative $\yt'$ is piecewise constant
($+1/T$, then $-1/T$, then $0$), jumping at each kink. The right-hand
data entering \eqref{eq:eom-nondim} are therefore bounded but
discontinuous, and the natural strategy is to integrate segment by
segment, restarting at each kink.

\subsection*{Exact piecewise-analytic solution}
\label{app:analytic}

Let $u \equiv \zt - \yt$ be the suspension deflection. Subtracting
$\yt''$ from both sides of \eqref{eq:eom-nondim}:
\begin{equation}
  u'' + 4\pi\zeta\,u' + (2\pi)^{2}\,u = -\,\yt''(\tau) .
\label{eq16}
\end{equation}
For the triangular pulse, $\yt'$ is piecewise constant, so $\yt''$ is a
sum of delta functions at the kinks:
\begin{equation}
  \yt''(\tau) = \sum_{k} \Delta_k\,\delta(\tau - \tau_k),
  \qquad
  (\tau_k;\ \Delta_k) =
  \Bigl(-T;\ \tfrac{1}{T}\Bigr),\
  \Bigl(0;\ -\tfrac{2}{T}\Bigr),\
  \Bigl(T;\ \tfrac{1}{T}\Bigr).
  \label{eq17}
\end{equation}
Each delta launches the impulse response of the homogeneous oscillator,
\begin{equation}
  G(s) = e^{-2\pi\zeta s}\,
         \frac{\sin\bigl(\omega_d\, s\bigr)}{\omega_d}\,
         \theta(s),
  \qquad
  \omega_d \equiv 2\pi\sqrt{1-\zeta^{2}} ,
\end{equation}
with $\theta$ the unit step, so that
\begin{equation}
  \zt(\tau)
  = \Lambda\!\left(\frac{\tau}{T}\right)
  \;-\; \frac{1}{T}\Bigl[\,G(\tau+T) - 2\,G(\tau) + G(\tau-T)\,\Bigr] .
  \label{eq:exact}
\end{equation}
The structure is transparent: the body follows the ramp exactly (a
linear $\yt$ is a steady state of \eqref{eq:eom-nondim}, since it makes
both the relative displacement and relative velocity terms cancel
against zero acceleration), and the entire dynamical response is a
\emph{second difference} of impulse responses fired at the three kinks
--- the discrete analogue of the fact that the oscillator responds to
$\yt''$. Differentiating twice (away from the kinks, where $\yt'' = 0$)
gives the body acceleration
\begin{equation}
  \zt''(\tau)
  = -\frac{1}{T}\Bigl[\,G''(\tau+T) - 2\,G''(\tau) + G''(\tau-T)\,\Bigr],
\end{equation}
where explicitly, for $s > 0$,
\begin{equation}
  G''(s) = e^{-2\pi\zeta s}
  \left[
    \frac{4\pi^{2}\zeta^{2}-\omega_d^{2}}{\omega_d}\,
      \sin(\omega_d s)
    \;-\; 4\pi\zeta\,\cos(\omega_d s)
  \right] .
\end{equation}
(The delta functions formally present in $G''$ at $s=0$ cancel the
$\yt''$ deltas exactly --- this is just the statement that the body acceleration takes finite jumps at the kinks.)
Expressions \eqref{eq:exact} and its derivatives can be evaluated in closed
form at any $\tau$ and any $(T,\zeta)$.

\begin{figure}[ht!]
\centering
\includegraphics[width=0.8\textwidth]{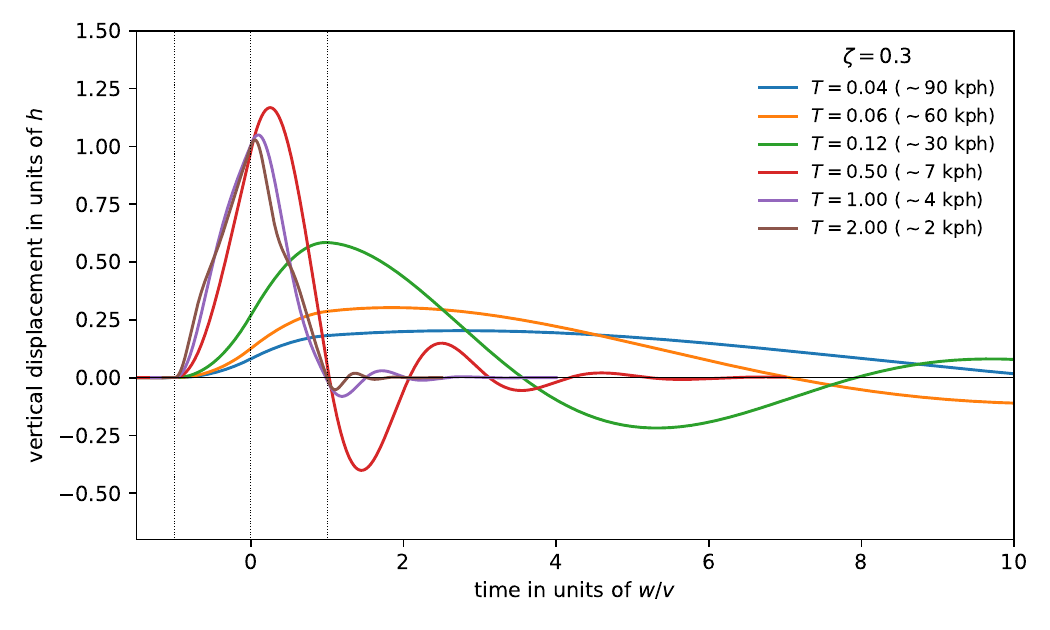}
\caption{Body vertical displacement $z$ (in units of bump size $h$) for various values of 
nondimensional $T$, parameterizing how fast the car crosses the bump.
Slow crossings ($T \gtrsim 1$) track the triangular
bump quasi-statically; in the range $T \sim 0.1$ to $0.5$ the
crossing resonates with the
suspension and the body; fast crossings
($T \lesssim 0.1$) leave the body increasingly unmoved.}
\label{fig:zhist}
\end{figure}

Figure \ref{fig:zhist} shows the solution of equations \eqref{eq16}--\eqref{eq17} for the vertical displacement, for various values of $T$, with $\zeta = 0.3$. The result is not surprising: For slow crossings of the bump, the suspension is quasi-statically rigid, and the body goes up and down in the shape of the bump. In the opposite limit of fast crossing, the bump is
over before the car can respond, and the displacement response
amplitude shrinks with $T$. In the intermediate range of
$T \sim 0.1$ to $0.5$, the suspension of the car is excited near resonance,
producing overshoots and ringing. The vertical dashed lines at $-1$, $0$, and $1$ mark the wheel entering, cresting, and exiting the ramp, respectively.

\begin{figure}[ht!]
\centering
\includegraphics[width=0.8\textwidth]{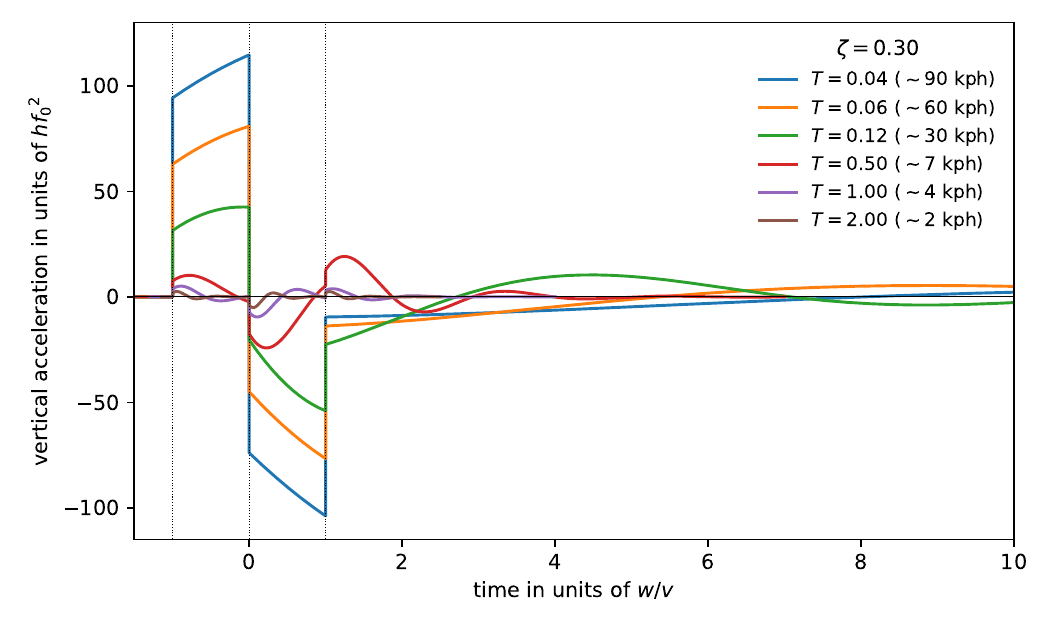}
\caption{Body acceleration $\ddot{z}$ in units of $hf_0^2$ for the same
cases as Figure~\ref{fig:zhist}. The acceleration is piecewise smooth
with finite jumps at the three kink times $\tau = -T,\,0,\,+T$. the
suspension converts the axle's impulsive accelerations into steps,
followed by damped ring-down at the suspension frequency. The jump
amplitudes grow as $1/T$: what the displacement response loses at
speed, the acceleration response gains.}
\label{fig:zddhist}
\end{figure}

Displacement is not,  however, what the driver directly feels. The human body responds, rather, to acceleration (or to some extent the derivative of acceleration, known as {\em jerk}). Figure \ref{fig:zddhist} shows the time course of the body acceleration for same values of $T$.

The main features of the figure are easily understood. Consider first the case of fast crossing, $T \ll 1$: At time $-1$ (in the units of the figure), the wheel hits the upgoing ramp, imposing an immediate, constant upward velocity on the wheel. The car's shock absorber (dashpot shown in Figure 2) thus imposes an immediate force, i.e., acceleration, on the body, acting through the term $2\zeta\omega_0\,(\dot z - \dot y)$
in equation \eqref{eq:eom-omega}. This is the step increase seen. For small $T$,
the body position is approximately unchanged during ramp climb (Figure 3),
so the spring is compressed approximately linearly with time, with acceleration
increasing up to an added acceleration of $\omega_0^2 h$ at the crest.

At time $0$, the upward slope changes to a downward slope, and the acceleration crashes down to a corresponding negative value. At time $1$ (ramp exit), the velocity of the wheel goes from negative to zero, producing (again via the shock-absorber) the third step-function change in the body acceleration. The other features seen in the figure are the suspension response.

For values of $T \sim 1$ (more obviously in the range 0.1 to 0.5), the bumps occur near resonance of the suspension. Since the speeds are slow, the accelerations are small, but one sees overshoot and ringing.

Now understanding Figure \ref{fig:zddhist}, one may reasonably ask, ``What happened to the Schwitters Doctrine?" Specifically, the faster we drive across the bump, the greater the accelerations, and sudden changes in acceleration (large jerk), that we experience. From the figure, one might conclude that we should slow to $T\gtrsim 1$, only a few kilometers per hour. In fact, many people annoyingly do!

\section{Schwitters Doctrine Found}
\label{sec:contact}

Sometimes, in physics problems of this sort, one can be misled by an over-reliance on formalization, missing physical effects that the formalism failed to incorporate. Here, the missed effect is the gravitational acceleration $g$.

For $T \ll 1$, the up-ramp imposes on wheel a vertical velocity $vh/w$, implying (via the shock absorber) an acceleration
\begin{equation}
    a_\text{ramp} = 2\zeta\omega_0\, vh/w
\end{equation}
for a time $w/v$. So, a ``universal" upward velocity
\begin{equation}
    u \equiv a_\text{ramp}\, w/v = 2\zeta\omega_0 h \approx 56\text{ cm/s}
\end{equation}
is imparted, notably independent of $v$. While this sudden impulse looks scary in Figure 4,
an equivalent statement is simply that the padded car seat unexpectedly collides with the driver's bottom with velocity $u$, the equivalent of a fall from a height
\begin{equation}
    d = \frac{u^2}{2g} \approx 1.6\,\text{cm.}
\label{eq:fall}
\end{equation}
Falling that far into a padded seat isn't much of an event! So a first important point is that one's feeling of the first bump is small and independent of how fast one is driving.

Now consider what happens at the crest (i.e., apex). The car body has
an upward velocity $u$, and the wheel on the ground now pulls down at velocity $vh/w$, implying, via the shock absorber, an upward force on the wheel. The spring, still compressed by $\approx h$, provides a downward force. When the net of these forces on the wheel exceeds the gravity $g$, the wheel lifts off the ground, a condition that we
can write as
\begin{equation}
    2\zeta\omega_0\!\left(u + \frac{vh}{w}\right)
    - \omega_0^{2} h  > g
\label{eq25}
\end{equation}
With some algebra, and using $2\zeta\omega_0 u = 4\zeta^{2}\omega_0^{2}h$, this condition becomes
\begin{equation}
    v > v_\text{crest} = v_\text{crit}
    \left[\,1 + (1-4\zeta^{2})\,\frac{\omega_0^{2}h}{g}\,\right]
    \approx 2400\,\text{cm/s} \approx 86\,\text{km/h} \approx 54\,\text{mph},
\end{equation}
where $v_\text{crit}$ is given simply by
\begin{equation}
    v_\text{crit} = \frac{gw}{u} \approx 1750 \text{ cm/s } \approx 63 
    \text{ km/h } \approx 40 \text{ mph}
\end{equation}
For reference in Figures 3 and 4, at $v=v_\text{crest}$, the dimensionless crossing time $T$ is
\begin{equation}
    T_\text{crest} = \frac{f_0 w}{v_\text{crest}}
    \approx 0.04
\label{eq:tcrit}
\end{equation}
(Note that $T_\text{crest} \ll 1$ for any velocity $v \gg f_0 w\approx 3.6$ km/h.)

For velocity $v > v_\text{crest}$, the wheel launches at the crest.
For velocity $v_\text{crest} > v > v_\text{crit}$, the wheel launches from the road somewhere on the downslope. To see this, note that two
quantities decay as the wheel descends: the spring compression,
$y - z \approx h(1 - x/w)$, linearly by construction; and the body's
upward velocity, which starts at $u$ and is consumed by the damper
deceleration $\approx 2\zeta\omega_0 hv/w$ acting over the slope time
$w/v$ --- a total decrement $2\zeta\omega_0 h = u$, independent of
$v$, so that $\dot z(x) \approx u(1 - x/w)$ as well. Two terms in
equation \eqref{eq25}
therefore relax with the same linear factor, while the remaining third term $2\zeta\omega_0vh/w$ remains constant.
From this, one can calculate that the launch point moves linearly,
\begin{equation}
  \frac{x_\text{launch}}{w}
  \;=\; 1 \;-\; \frac{v - v_\text{crit}}{v_\text{crest} - v_\text{crit}} ,
\end{equation}
sliding from the base of the slope at $v = v_\text{crit}$ up to the
crest at $v = v_\text{crest}$.

When the wheel and body are launched into freefall, the body is still approximately at its flat-road equilibrium position. On a parabolic trajectory, it returns to that height after a horizontal distance
\begin{equation}
    d_\text{landing} = \frac{2vu}{g} = 2 w\frac{v}{v_\text{crit}} \approx 2.8 w\frac{v}{v_\text{crest}}
\label{eq:dlanding}
\end{equation}
which puts the landing at least a distance $w$ beyond the end of the ramp down-slope (more at higher speed). Actually, we will see in \S\ref{sec:limitations} that equation \eqref{eq:dlanding} is a lower bound: The body's parabolic trajectory on the way down is slowed by meeting the shock absorber and settling to the equilibrium position only slowly and farther from the bump.

To summarize, at the crest, when $v > v_\text{crest}$, the driver feels no bump at all (road pulls downward away from car), a period of reduced weight lasting $2u/g \approx 0.12$ sec, and then a shock-absorber softened landing on smooth road.
The landing bump is the car body falling onto its suspension with velocity $u$, producing a peak deceleration (via the shock absorber) of
\begin{equation}
    a_\text{landing} = 2\zeta\omega_0 u = \frac{u^2}{h} \approx 0.22 g
\label{eq:alanding}
\end{equation}
and lasting for a time
\begin{equation}
    \tau_\text{ landing}  = \frac{2h}{u} \approx 0.5\,\text{s}
\end{equation}
both the acceleration and its duration independent of the velocity $v$.

We recover the Schwitters Doctrine: At any velocity $v > v_\text{crest}$, the driver experiences the onset of the ramp as the equivalent of a 1.6 cm fall into the seat, a $\approx 0.12$ s period of reduced weight, and a 0.2 g landing deceleration lasting about 0.5 s. In the model (and ignoring safety concerns!) there is no reason not to take the speed bump as fast as possible.

\section{Limitations of the model}
\label{sec:limitations}

The model treated above is highly idealized: a single
massless wheel, rigid tire, linear spring and damper, representing one quarter of a
vehicle. Each idealization deserves scrutiny. We show here that none
overturns the Schwitters Doctrine. Several modify the numbers somewhat, while one (the
wheel's own flight) adds a phenomenon of its own.

\subsection*{Effect of pneumatic tires}

Let $m_w$ denote the mass of the wheel and associated unsprung components,
with typically $m_w \sim 0.1 \,M$. The weight of a car compresses its pneumatic tires by typically $\sim 2$ cm,
as compared to the static sag on its suspension of $\sim 20$ cm, implying a
tire spring constant $k_\text{tire} \sim 10\,k$. Tire and wheel form a classic high-frequency isolator, mitigating vibrations at frequencies larger than the corner frequency
\begin{equation}
    f_\text{tire} = \frac{1}{2\pi} \sqrt{\frac{k_t}{m_w}} \approx \frac{1}{2\pi} \sqrt{\frac{10 k}{0.1 M}} = 10 f_0 \approx 10 \text{ Hz}
\end{equation}

Above the corner frequency the tire deflects instead of the shock absorber
delivering impulse. One consequence is that the crest
velocity-reversal spike ($\sim 2rh/wv$, a few ms at speed) is absorbed, and
the tire, not the suspension, is what
keeps real wheels on the downslope at speeds $<\!v_{\text{crit}}$.
Another is that the tire contact patch envelopes small road irregularities shorter than its length, $\sim\!10$~cm.

\subsection*{Frame of the car}

The quarter-car model treats each corner separately, while in reality the
corners are joined by a stiff frame. Perhaps by design, the length of the speed bump $2w$ is not too different from the wheelbase $L$ of a typical car, tending
to maximize the front-rear pitch angle produced in the quasi-static case $T \gtrsim 1$.
On the other hand, since more typically we have  $T \ll 1$ (equation \eqref{eq:tcrit}), the front and rear impulses are delivered to the car's suspension near-simultaneously
on its timescale $f_0$, mitigating the pitch. If only one side of the axle (left or right) encounters the bump, there is no such compensation, and the torsional stiffness of the suspension comes into play,
beyond our scope here.

\begin{figure}[ht!]
\centering
\includegraphics[width=0.8\textwidth]{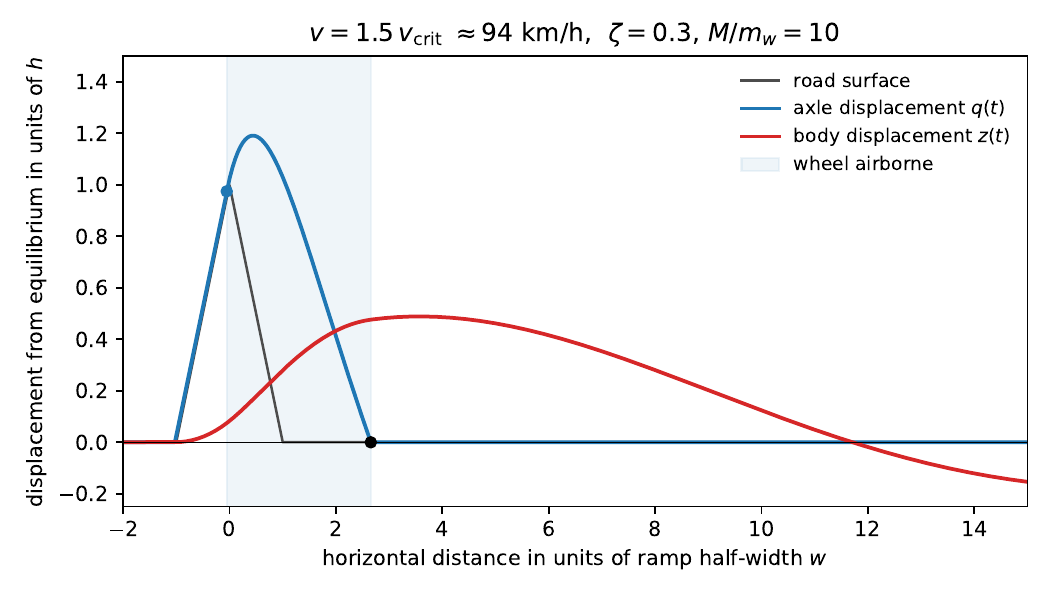}
\caption{Numerical integration of the two-body (body and wheel) ODE equations of motion. With $v > v_\text{crest}$, the wheel (blue curve) leaves the road, putting the system center of mass into free fall. The body (red curve) approximates a parabolic trajectory until the wheel returns to the ground, anchoring the shock absorber. After that, the body settles onto its $\sim\!\!1$ Hz suspension.}
\label{fig:numerical}
\end{figure}

\subsection*{Trajectory of the wheel}

For finite $m_w$, it is the center of mass of body plus wheel that follows
the free-fall parabola, while body and wheel individually exchange
momentum through the suspension. The body's acceleration during flight is
$\ddot z = [\,k \xi + c\dot \xi\,]/M$, with $\xi$ the spring extension, and
reaches $-g$ only as the spring relaxes toward equilibrium.
The extension $\xi$
satisfies an equation of motion
\begin{equation}
    \mu\ddot\xi + c\dot\xi + k\xi = 0
\end{equation}
with $\mu$ the reduced mass
\begin{equation}
    \mu \equiv \left(\frac{1}{m_w} + \frac{1}{M}\right)^{-1}
\end{equation}
One easily derives (cf.~equations \eqref{eq:eom-dim}--\eqref{eq:eom-omega})
\begin{equation}
f_i = f_0\sqrt{\frac{M}{m_w}} \approx 3\,\text{Hz}, \qquad \zeta_i = \zeta\sqrt{\frac{M}{m_w}} \approx 0.95
\end{equation}
(numerical results with $M/m_w \approx 10$), where subscript $i$ here denotes the ``internal" degree of freedom $\xi$. Since $\zeta_i$ is nearly
critically damped, the strut extends monotonically, without bounce,
over a few tens of milliseconds. The driver therefore enters
weightlessness through a smooth $\sim 30$--$50$~ms ramp rather than
instantaneously, and the ``free fall'' of the previous section is the
body's motion only after this transient; corrections to the parabola
are $O(m_w/M) \approx 10\%$.

The wheel leaves the crest with the ramp upward velocity $vh/w$, but it is quickly hauled downward by the spring preload $Mg$ acting on the small mass $m_w$, a huge effective deceleration $\sim g \, M/m_w$. Its ascent is fast replaced by downward motion at the terminal damper-limited rate
\begin{equation}
  \dot \xi_{\mathrm{term}} \approx \frac{Mg}{c}
  = \frac{g}{2\zeta\omega_0} \approx 260~\mathrm{cm/s},
\end{equation}
taking a time
\begin{equation}
    \tau_\text{wheel} = h/\dot \xi_{\mathrm{term}} = hc/Mg = u/g \approx 0.06 ~\text{s}
\end{equation}
to descend a ramp height $h$. During that time, the horizontal distance moved is
\begin{equation}
    d_\text{wheel} = \frac{vu}{g} = w\,\frac{v}{v_{\mathrm{crit}}} \approx 1.4  w\,\frac{v}{v_{\mathrm{crest}}}
\label{eq:dwheel}
\end{equation}
which, for $v > v_{\mathrm{crit}}$, is beyond the end of the ramp.

Thus, the wheel is down, parked, and settled
before the body descends onto it, which is what justified
treating the landing as a body onto a stationary suspension, the
$\approx 0.22\,g$ damper catch, equation \eqref{eq:alanding}; and which
made equation \eqref{eq:dlanding} above (parabolic flight touchdown) only a lower bound.
The wheel's own touchdown, at a
closing speed of order $\dot \xi_{\mathrm{term}}$, is absorbed inelastically by the unsprung mass and is not directly felt by the driver.

Figure 5 shows the result of a numerical integration of the full coupled ODEs for wheel and body. The assumed velocity is $1.5\, v_\text{crit}$ (also exceeding $v_\text{crest}$). One sees the wheel's takeoff at the crest and landing at about $2.6 \,w$ beyond the crest (cf.~equation \eqref{eq:dwheel}'s underestimate). While the wheel is in the air, the body approximates a parabolic trajectory, but its upward path is arrested, and its downward path much softened by the now-grounded wheel.

\subsection*{Other effects}
Real shock absorbers are deliberately nonlinear. Blow-off valves clip forces
at high piston speed, softening the high-acceleration ramp-entry spike.
Suspensions have finite travel, with droop stops that end the strut's
relaxation before full extension and bump stops that stiffen large
compressions.

The seat cushion, a further $\sim\!2$--$3$~Hz stage in
series, regularizes the body's acceleration steps as ramps of finite jerk before they reach the driver. (We described the effect of the seat in a different way, analogous to a fall from $\approx\! 1.6$ cm height, in equation \eqref{eq:fall}, above.)

All of these effects act in the same direction: they
soften the sharp features that the rigid linear model overstates,
while leaving the slow, large-amplitude physics---the universal
launch velocity $u$, the reduced weight interlude, the gentle spring-borne landing---untouched. The Schwitters Doctrine appears to be more robust in a
real car than in the model here. Only sympathy for the mechanical and structural components of the suspension and frame, and a reasonable concern for safety, argue for restraint.

\section{Related work}
\label{sec:related}

The folk practice that a rough feature is best taken fast is
well established in dirt-road driving lore, where it is known as the corrugation sweet spot, typically quoted as 50--70~km/h \cite{Engineer,Windsor,Washboard}. At the sweet spot,
the wheels skim the ripple crests rather than tracking each trough.

The scientific basis for that practice is the literature on washboard road formation.  Mather~\cite{Mather1963}, in rotating-table experiments
with a spring-mounted wheel on sand, established that above a critical
speed a rolling wheel ceases to follow the surface and advances in a
sequence of short hops. By his account, this is both the reason that faster travel is
smoother and the mechanism by which corrugations form.
The periodic landings excavate the pattern that subsequent traffic
reinforces. Our $v_{\mathrm{crit}}$ is a suspension-borne, single-bump
cousin of Mather's critical speed: in both cases a unilateral contact
fails when the surface recedes faster than the restoring forces can
press the wheel after it, and the comfortable speed and the
road-damaging speed are the same.

The modern granular-physics literature has made washboard formation
quantitative. Taberlet, Morris, and McElwaine~\cite{Taberlet2007}
reproduced the instability in controlled laboratory experiments and
soft-particle simulations, locating a critical speed above which
ripples grow from an initially flat sand bed; the follow-up
study~\cite{Bitbol2009} organized the onset by a Froude-like
dimensionless group balancing dynamic against gravitational-frictional
forces---the same competition, $v^{2}$ versus $g\,\ell$ for the
relevant length $\ell$, that underlies our contact criteria.
Analytical models of the coupled wheel-surface instability go back to
Both, Hong, and Kurtze~\cite{Both2001}, and Hewitt, Balmforth, and
McElwaine~\cite{Hewitt2012} showed that even a rigid object dragged
over a \emph{fluid} surface washboards, demonstrating that the
instability requires neither granularity nor suspension dynamics.

Our problem is the complementary, deliberately simpler one: the road
profile is prescribed and rigid, all the dynamics live in the vehicle,
and much can be carried to closed form.

\section*{Acknowledgment and Disclosure}

Schwitters was an outstanding experimental physicist with whom I interacted
as a colleague, first at Harvard and later at The University of Texas at Austin. I remain grateful for these interactions, in which I learned a lot of real-life physics.

One big change since Roy's death is the ability of Large Language Models to solve quite complicated physics problems at undergraduate level, and to explain their solutions in pedagogically appropriate language. This paper embodies a fruitful collaboration with Claude Fable 5 \cite{Fable5,SystemCard}. Roy would be amazed and pleased at Fable's abilities. However, the physics-problem equivalent of so-called AI slop \cite{wikislop} is Fable's tendency, when called out on a poor explanation, to double down with longer and more baroque elaborations. That is where the human partner needs to understand the physics and not just be a cheerleader to the LLM. I have checked Fable's work, eliminated (I hope) any physics slop, and am responsible for any errors that remain.

I am grateful to Doug Finkbeiner, Jonathan Katz, Dan Meiron, and Richard Price for comments on the manuscript.

\bibliographystyle{unsrturl}
\bibliography{sample}

@article{Riordan2023,
  author  = {Riordan, Michael},
  title   = {{Roy Frederick Schwitters}},
  journal = {Physics Today},
  volume  = {76},
  number  = {4},
  pages   = {60--61},
  month   = apr,
  year    = {2023},
  doi     = {10.1063/PT.3.5224},
  note    = {Obituary}
}

@online{airhart2023,
  author       = {Airhart, Marc},
  title        = {{Remembering High-Energy Physicist Roy Schwitters}},
  year         = {2023},
  month        = jan,
  day          = {18},
  organization = {College of Natural Sciences, The University of Texas at Austin},
  url          = {https://cns.utexas.edu/news/announcements/remembering-high-energy-physicist-roy-schwitters},
  urldate      = {2026-07-10}
}

@online{wikiroy,
  author       = {{Wikipedia contributors}},
  title        = {{Roy Schwitters}},
  year         = {2026},
  organization = {Wikipedia, The Free Encyclopedia},
  url          = {https://en.wikipedia.org/w/index.php?title=Roy_Schwitters&oldid=1341073324},
  urldate      = {2026-07-10},
  note         = {Last edited 1 March 2026}
}

@online{mathworks,
  author       = {{The MathWorks, Inc.}},
  title        = {The Physics of the Damped Harmonic Oscillator},
  organization = {MathWorks},
  url          = {https://www.mathworks.com/help/symbolic/physics-damped-harmonic-oscillator.html},
  urldate      = {2026-07-11},
  note         = {Symbolic Math Toolbox example}
}

@online{openstax,
  author       = {{Physics LibreTexts}},
  title    = {{University Physics I. Mechanics, Sound, Oscillations, and Waves}},
  organization = {OpenStax},
  date         = {2022-09-12},
  url          = {https://phys.libretexts.org/Bookshelves/University_Physics/University_Physics_(OpenStax)/Book%3A_University_Physics_I_-_Mechanics_Sound_Oscillations_and_Waves_(OpenStax)},
  urldate      = {2026-07-11},
  note         = {See Section 15.6, ``Damped Oscillations"}
}

@book{young_freedman,
  author    = {Young, Hugh D. and Freedman, Roger A.},
  title     = {Sears and Zemansky's University Physics},
  edition   = {13},
  publisher = {Addison-Wesley},
  address   = {San Francisco, CA},
  year      = {2012},
  isbn      = {978-0-321-69689-2},
  note      = {See Sections 14.7, ``Damped Oscillations,'' and
               14.8, ``Forced Oscillations and Resonance''}
}

@misc{Engineer,
  author       = {{Engineer Fix}},
  title        = {How to Drive Safely on Washboard Roads},
  howpublished = {\url{https://engineerfix.com/how-to-drive-safely-on-washboard-roads/}},
  note         = {Accessed July 2026}
}

@misc{Windsor,
  author       = {{Windsor RV}},
  title        = {Driving Tips for {RVs} on Corrugated Roads},
  howpublished = {\url{https://www.windsorrvs.com.au/blog/driving-tips-corrugated-roads}},
  note         = {Accessed July 2026}
}

@misc{Washboard,
  author       = {{iRV2 Forums}},
  title        = {Washboard Dirt Roads: Is There a Magic Speed?},
  howpublished = {\url{https://www.irv2.com/forums/f50/washboard-dirt-roads-is-there-a-magic-speed-587656.html}},
  year         = {2022},
  note         = {Accessed July 2026}
}

@article{Mather1963,
  author  = {Mather, Keith B.},
  title   = {{Why Do Roads Corrugate?}},
  journal = {Scientific American},
  volume  = {208},
  number  = {1},
  pages   = {128--136},
  month   = jan,
  year    = {1963},
  doi     = {10.1038/scientificamerican0163-128}
}

@article{Taberlet2007,
  author  = {Taberlet, Nicolas and Morris, Stephen W. and McElwaine, Jim N.},
  title   = {Washboard Road: The Dynamics of Granular Ripples Formed by
             Rolling Wheels},
  journal = {Physical Review Letters},
  volume  = {99},
  pages   = {068003},
  year    = {2007},
  doi     = {10.1103/PhysRevLett.99.068003}
}

@article{Bitbol2009,
  author  = {Bitbol, Anne-Florence and Taberlet, Nicolas and
             Morris, Stephen W. and McElwaine, Jim N.},
  title   = {Scaling and dynamics of washboard roads},
  journal = {Physical Review E},
  volume  = {79},
  pages   = {061308},
  year    = {2009},
  doi     = {10.1103/PhysRevE.79.061308}
}

@article{Both2001,
  author  = {Both, Joseph A. and Hong, Daniel C. and Kurtze, Douglas A.},
  title   = {Corrugation of roads},
  journal = {Physica A: Statistical Mechanics and its Applications},
  volume  = {301},
  number  = {1--4},
  pages   = {545--559},
  year    = {2001},
  doi     = {10.1016/S0378-4371(01)00425-3}
}

@article{Hewitt2012,
  author  = {Hewitt, Ian J. and Balmforth, Neil J. and McElwaine, Jim N.},
  title   = {Granular and fluid washboards},
  journal = {Journal of Fluid Mechanics},
  volume  = {692},
  pages   = {446--463},
  year    = {2012},
  doi     = {10.1017/jfm.2011.523}
}

@online{wikislop,
  author       = {{Wikipedia contributors}},
  title        = {{AI slop}},
  year         = {2026},
  organization = {Wikipedia, The Free Encyclopedia},
  url          = {https://en.wikipedia.org/wiki/AI_slop},
  urldate      = {2026-07-10},
}

@misc{Fable5,
  author       = {{Anthropic}},
  title        = {Introducing {Claude} {Fable} 5 and {Claude} {Mythos} 5},
  howpublished = {\url{https://www.anthropic.com/news/claude-fable-5-mythos-5}},
  month        = jun,
  year         = {2026},
  note         = {Accessed July 2026}
}

@misc{SystemCard,
  author       = {{Anthropic}},
  title        = {Claude {Fable} 5 and {Claude} {Mythos} 5 System Card},
  howpublished = {\url{https://www-cdn.anthropic.com/d00db56fa754a1b115b6dd7cb2e3c342ee809620.pdf}},
  month        = jun,
  year         = {2026},
  note         = {Accessed July 2026}
}

\end{document}